\documentclass[aps,prl,twocolumn,superscriptaddress,showpacs,floatfix]{revtex4}
\usepackage{graphicx}  
\usepackage{dcolumn}   
\usepackage{bm}        
\usepackage{amssymb}   

\begin{document}


\title{Anomalous Noise in the Pseudogap Regime of 
YBa$_2$Cu$_3$O$_{7-\delta}$}

\author{D. S. Caplan}
\author{V. Orlyanchik}
\email{vlor@illinois.edu}
\author{M. B. Weissman}
\author{D. J. Van Harlingen}
\author{E. H. Fradkin}

\affiliation{Department of Physics and Materials
Research Laboratory, University of Illinois at
Urbana-Champaign, Urbana, Illinois 61801, USA}
\author{M. J. Hinton}
\author{T. R. Lemberger}
\affiliation{Department of Physics, Ohio state
University, Columbus, Ohio 43210, USA}

\date{\today}

\begin{abstract}
An unusual noise component is found near and below about 250 K in the normal state of underdoped 
YBCO and Ca-YBCO films. This noise regime, unlike the more typical noise above 250 K, has features expected for a symmetry-breaking collective electronic state. These include large individual fluctuators, a magnetic sensitivity, and aging effects. A possible interpretation in terms of fluctuating charge nematic order is presented.

\end{abstract}


\maketitle

Key questions about pseudogap phenomena \cite{Timusk99,Norman05} in high-$T_c$ 
superconductors remain unsettled.  Some descriptions of pseudogap physics involve  at least local breaking of various symmetries, including translational invariance, point group symmetries 
and time-reversal \cite{kivelson-1998,chakravarty-2001c,Kivelson03,varma-2005,Lee06,Vojta09}. 
Recently experimental evidence for such ordered states has been found in many cuprates, including the cleanest YBa$_2$Cu$_3$O$_{7-\delta}$ (YBCO) materials. STM images in BSCCO\cite{Howald03,kohsaka-2007} show local  electronic nematic order. Static stripes are shown by neutron scattering\cite{tranquada-2007} in LBCO near $x=1/8$ and in the LSCO family at finite magnetic fields\cite{lake-2002}, and by doping-dependent quantum oscillations in YBCO\cite{doiron-leyraud-2007}. Hysteretic magnetotransport has been found in LSCO samples\cite{Popovic}. Time reversal breaking has been seen in YBCO by inelastic neutron scattering (INS)\cite{Fauque06,Mook08} and by the Kerr effect\cite{Kapitulnik08}. Nematic order has also been detected in very underdoped YBCO by INS\cite{hinkov-2007} and in transport\cite{Ando02}, including the Nernst effect\cite{taillefer-nematic-2009}. Partially static order is suggested by hysteretic effects seen in YBCO in magnetic measurements \cite{Panag04,Panag06,Kapitulnik08,Sonier09} 
and in transport\cite{Bonetti04}.

\begin{figure}[h]
\centering
\includegraphics[width=0.45\textwidth]{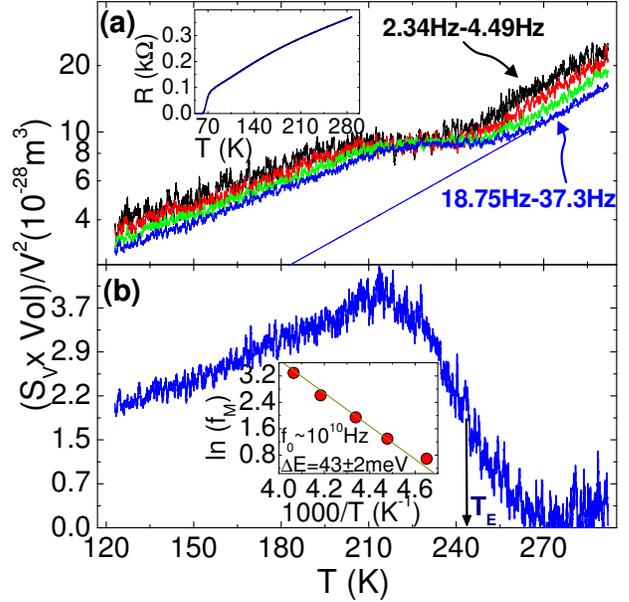}
\caption{\label{Tpeak} (color online)  (a) Noise power, S$_V$, in four consecutive octaves vs. T smoothed over the range of 0.3 K for the sample with T$_c$=58 K. Note the significant deviation of the noise power from what would be expected by extrapolating from high T part (solid line). Inset: Dependence of the resistance on temperature; (b) Excess noise after subtraction of high-T part (see text) shown by a straight line in (a) for octave 6. The arrow marks T$_E(f)$ (see text). Inset: Dependence of T$_E$ on measurement frequency. Solid line is Arrhenius fit to the data.}
\end{figure}

It remains unclear whether the transport anisotropy effects in untwinned crystals of  YBa$_2$Cu$_3$O$_{7-\delta}$ (YBCO), attributed to stripes \cite{Ando02}, are related to the peculiar magnetic memory effects \cite{Kapitulnik08}. If the transport anisotropy comes from stripes, then any slow fluctuations in this local symmetry breaking should give rise to transport noise \cite{Carlson06} as found in diverse other systems including antiferromagnets \cite{Israeloff88}, spin glasses \cite{Israeloff89}, and ferromagnets \cite{Merithew00}. Discrete resistance steps seen in small samples of YBCO \cite{Bonetti04}, and an unusual increase in normalized noise power found in larger samples \cite{Bei97,Bei00} also suggest large collective fluctuators, as expected from stripes. So far properties such as distinctive temperature dependence or magnetic sensitivity needed to clearly connect such noise with stripe-like physics have not been reported.

In this Letter we report that low-frequency transport noise in underdoped YBCO and Ca-doped Y$_{1-x}$Ca$_x$Ba$_2$Cu$_3$O$_{7-\delta}$ (Ca-YBCO) films shows a well defined  temperature, around 250 K, dependent on measurement frequency but very little on doping, below which an extra noise component, exhibiting magnetic sensitivity, aging effects and some large discrete fluctuators, sets in. Such effects are characteristic of noise from collective states in disordered systems \cite{Weissman88, Israeloff88, Israeloff89, Merithew00}. Magnetic memory effects in the transport noise implies its connection with magnetic hysteresis \cite{Kapitulnik08}, suggesting coupling to some other magnetic order. The large fluctuators allow for mesoscopic thermodynamic characterization of the states involved. \cite{Weissman88, Merithew00}. Noise symmetry measurements \cite{Weissman88} further characterize the collective order.

Thin YBCO and Ca-YBCO films (30-35 nm) were grown by pulsed laser deposition. Except for one film grown on a vicinal SrTiO$_3$ (001) (STO) substrate to reduce twinning \cite{Hilgenkamp03}, LaAlO$_3$ (LAO) substrates were chosen for their lower low-frequency noise \cite{Rajeswari96}. 
The films were first nearly optimally doped, then annealed in low O$_2$ pressure to obtain 30K$<$T$_c<$85K corresponding to 0.65$>\delta>$0.3. Samples were configured by photolithography and ion milling for four-probe noise measurements, except for two eight-probe samples used for symmetry measurements \cite{Weissman88}. Typical samples were 2-3 $\mu$m wide and 15 $\mu$m long. The contact area was ion milled before depositing the gold contacts to minimize their noise. Resistance (R) vs. temperature (T) showed typical behavior for underdoped material, as in the inset to Fig.~\ref{Tpeak}(a). The partially untwinned sample showed a typical resistive anisotropy ratio between the two inequivalent crystal directions \cite{Ando02}.

To measure noise a DC current was passed through the sample and the resulting voltage drop was fed into an AC coupled SR552 low noise preamplifier followed by an anti-alias filter SR640. We compute the voltage spectral density, S$(f,T)$,  for 0.3 Hz$<f<$112 Hz from the filtered AC voltage 
digitized at 300 Hz. A zero-current background spectrum is subtracted to obtain S$_V(f,T)$, the portion of S$(f,T)$ due to fluctuations in R. (S$_V(f,T)$ is quadratic in applied current 
up to the largest currents used.)  We compact and average S$_V(f,T)$ by integrating over f into octave sums, convenient for spectra of the form S$(f) \propto1/f^{\alpha}$, since for $\alpha=1$ each octave has the same power \cite{Weissman88}. During T-dependence measurements, T was swept continuously with typical rate of 0.1-0.3 K/min. No difference between data taken on cooling 
and heating was found.

Figure \ref{Tpeak}(a) shows the T-dependent octave sums (normalized by $V^2$) for the sample with a T$_c$ of 58 K. An extra low-temperature noise source appears to be present below about 250 K. 
We define T$_E(f)$ as T at which the excess of S$_V$(f) above its high-T extrapolation reaches half its maximal value. The f-dependence of T$_E(f)$ (see inset Fig.~\ref{Tpeak}(b)) follows thermally activated kinetics $f_M=f_0exp[-\Delta E/k_BT_E]$ ($f_M$ measurement frequency), with a typical 
attempt rate of $f_0\sim 10^{10}-10^{11}$ Hz and an activation energy of $\Delta E\sim 0.4$ eV. Thus, T$_E(f)$ gives no indication of a genuine phase or glass transition in this vicinity, just a rather sharp change in a distribution of activation energies.

This feature appeared in the noise power for all 13 samples measured, patterned from 7 separate thin films. Except for one sample from a film which showed several signs of significant inhomogeneity, T$_E(f)$ and the derived $\Delta E$ were nearly the same in samples with T$_c$'s 
ranging from 35 K to 85 K, regardless of substrate or Ca content. We shall later discuss the sharp difference between such behavior and measures of the onset of pseudogap correlations. As we now discuss, aging effects, large discrete fluctuators, and magnetic sensitivity found 
exclusively below $\sim$250 K indicate that the extra noise is due to large-scale collective effects associated with electronic symmetry breaking.

\begin{figure}
\includegraphics[width=0.45\textwidth]{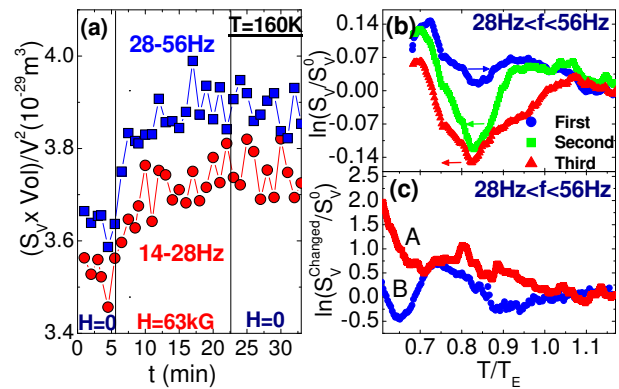}
\caption{\label{magnetic}(color online)  (a) Noise power before, during and after application of 6.3 T (H$||c$). (b) Log of ratio of noise power to pre-field value immediately after removal of H (dT/dt$\textgreater$0) and during 2 consecutive thermal cycles (dT/dt$\textless$0). Ca-YBCO, smoothed over the range of 3 K. Sample T$_C$=62 K. (c) Persistent changes in S$_V$ that occurred in two samples (sample A and B) from the same film with T$_C$=85 K. Significant changes in (b) and (c) are found only below T$_E$.}
\end{figure}

Fig.~\ref{magnetic}(a) shows an example of the magnetic sensitivity of the noise at low T, in this case an increase in magnitude of about 6$\%$ upon application of 6.3 T with H$\parallel $c.  Removing H gave no significant immediate change, but upon subsequent T-cycling the spectrum relaxed toward a different value, as shown in Fig.~\ref{magnetic}(b). Importantly, the changes in the spectrum, both in response to the field and over time, are negligible above $\sim$250 K.  Changes in spectra of other samples induced by magnetic fields differ in detail but also are negligible above $\sim$250 K.

As in the Kerr effect \cite{Kapitulnik08}, the noise memory of the magnetic perturbations persists even after cycling to room temperature. Unless YBCO has two different types of high-T magnetic memory, (the origin of even one such memory is currently unknown) the transport noise is connected with the same time-reversal symmetry-breaking seen via the Kerr effect. Both the noise and Kerr memories may involve special magnetic regions with high melting temperature, e.g. surface antiferromagnetism. However, neither method shows symptoms of these magnetic effects until T is 
lowered enough for more pervasive magnetic correlations to show up.

In another pair of samples, T$_c$=85 K, a persistent change in the T-dependent spectra occurred, only below  $\sim$250 K, in between  T-cycles (see Fig.~\ref{magnetic}(c)). The magnetic sensitivity of the noise (found below 250 K) also changed significantly. We found no accompanying change in R.  

The sensitivity of the noise to magnetic field changes, found only in the low-T regime, strongly suggests that this noise mechanism is associated with some sort of magnetic order. However, the results on the magnetic sensitivity of different samples to different magnitudes, histories, and orientations of fields are too complex for us to characterize without further study.

\begin{figure}
\centering
\includegraphics[width=0.4\textwidth]{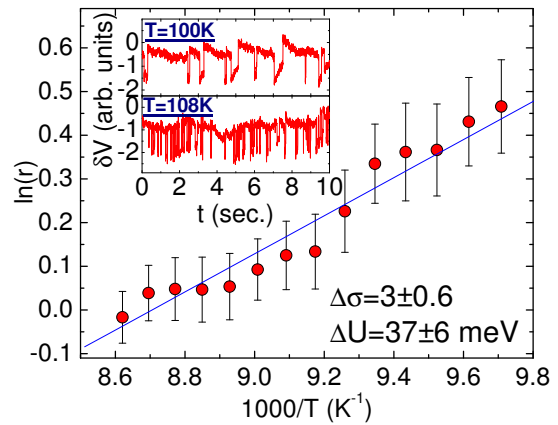}
\caption{\label{fluctuator}(color online)  Boltzmann factor for the fluctuator from the inset vs. 
inverse temperature. Solid line is a fit to a Boltzmann expression. Inset: Discrete two-level fluctuator at two temperatures. Sample: YBCO, T$_C$=85 K.}
\end{figure}

Below 200 K we often find large, T-dependent individual fluctuators (see Fig.~\ref{fluctuator}), mostly of the two-level form, showing activated kinetics with attempt rates of about $10^{10}$ Hz,  for which our frequency window limits observable  $\delta E$ to about 20k$_B$T. In all 13 samples, many on multiple T-sweeps, we found no discrete fluctuators above 200 K. These detectable discrete fluctuators have magnitudes $\delta R /R > 10^{-5}$, at least 3 orders of magnitude larger than would be expected due to changes in scattering from the motion of small defects if the 
conductivity were approximately uniform \cite{Weissman88}. Thus either the conductivity is highly non-uniform or there are large collective fluctuations, or both \cite{Merithew00}. Both non-uniformity and large fluctuators are expected in stripe-like pictures \cite{Carlson06}.

For two-level fluctuators we calculate the free energy difference between the levels, $\Delta F$, from the ratio $r(T)$ of time spent in each level from the Boltzmann expression $r(T)=exp(\Delta F / k_B T)$. The temperature derivative of $\Delta F$ then yields the change in entropy $\Delta \sigma$ and the change in energy $\Delta U$, using standard thermodynamic relations. Pure switching between different versions of a broken-symmetry phase should give 
zero for $\Delta \sigma$ and $\Delta U$. We find typical values of $|\Delta \sigma|<10$, but some outside the error bars from zero. We measured a fluctuation isotropy parameter \cite{Weissman88} ${S\equiv 2 \frac{\langle Det( \delta \boldmath{\rho})\rangle}{\langle Tr [(\delta  \boldmath{\rho})^2]\rangle}}$, where {\boldmath $\rho$} is the two-dimensional resistivity \cite{Weissman88}, in one YBCO sample with T$_c$=65 K. We found $S \sim$ 0, indicating that $\delta \rho$ is not a scalar ($S=1$), but also is not a rotation of an easy axis in an otherwise symmetrical environment ($S=-1$).

\begin{figure}[ht]
\includegraphics[width=0.4\textwidth]{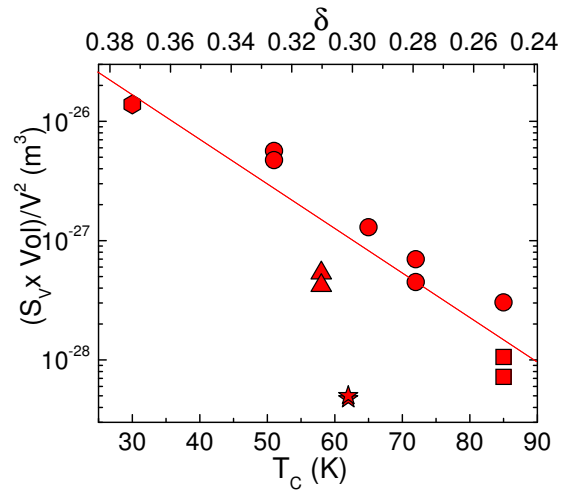}
\caption{\label{PeakT} (color online) Amplitude of the excess noise at T$_E(p)$ vs. T$_c$ and $\delta$ (oxygen deficiency) for each distinct sample. Circles and triangles are YBCO on LAO substrate 4 and 8 - terminal geometry, respectively. Stars and hexagon are Ca-YBCO on LAO substrate with 4 and 8 terminal geometry, respectively. Squares are YBCO on STO substrate with 4 
terminal geometry. The solid line is guide to the eyes.}
\end{figure}

If the noise mechanism involves resistive anisotropy associated with stripes \cite{Ando02}, it should be enhanced at low T and at low doping \cite{Ando02}. The largest excess noise is indeed  found in the more underdoped samples, as shown in Figure \ref{PeakT}, and the reduction of noise at low T is less than expected from purely kinetic effects given the spectral slope \cite{Weissman88}.  

In conclusion, below $\sim$250K an extra low-frequency noise source with qualitatively distinct features sets in. It  shows aging effects and individual fluctuators much larger than expected for any standard defect noise mechanism. These features are expected when some broken symmetry is partially pinned by disorder, giving quasi-static fluctuations.  The onset of some complicated magnetic field sensitivity of the noise in this low-T regime points toward a connection with other electronic effects associated with the pseudogap, rather than e.g. the ferroelastic order suggested to account for some low-frequency internal friction features \cite{Zheng02}. The recent Nernst effect evidence \cite{taillefer-nematic-2009}  for nematic order in the pseudogap temperature range over a broad doping range in YBCO samples also supports interpreting the  fluctuators as local nematic patches. 

In the simplest picture of nematic order \cite{Carlson06}, twinning provides a random field breaking up the simple ordered state, allowing noise from rotations of the easy axis of stripe domains. To the extent that nematic order is coherent over distances large compared to the twinning scale, the rotations would occur in symmetrical environments and hence not change scalar quantities. We find some fluctuations in scalar quantities: energy, entropy, and resistivity tensor trace. Thus the collective state is likely not to show rigid nematic order over distances much larger than the typical twinning scale of several tens of nanometers, consistent with neutron scattering results. \cite{haug-2009}

Although noise from disordered collective states is common, the onset of the low-T noise in these YBCO films is peculiar in showing simple thermally activated kinetics, rather than any sort of sharp transition or crossover \cite{Israeloff88, Merithew00, Weissman88}. Another low-frequency phenomenon in YBCO, internal friction, shows thermally activated features in this approximate temperature range \cite{Cannelli92, Zheng02}, but none of the features provide an obvious match to our noise results, and the origins of the friction peaks are themselves unclear. T$_E(f)$ lacks the doping dependence of signatures of the onset of electronic correlations (e.g. by neutron scattering and the Kerr effect \cite{Fauque06,Mook08,Kapitulnik08}). 

We are then left with a puzzle. Below T$_E(f)$ noise appears with the qualitative features expected from disordered correlated electronic states involving magnetism. The Arrhenius dynamical crossover into this regime is not itself the onset of dynamical cooperativity, which would show sharper T-dependence. The most important surprising feature of the noise is that it 'remembers' the effects of magnetic fields even after warming to room temperature, just as does the Kerr effect in  similar films \cite{Kapitulnik08}. There must be some magnetic order, e.g. antiferromagnetic surface layers, which remains even at such temperatures. That suggests a tentative explanation for the crossover into the low-temperature noise regime. Magnetic exchange interactions in these materials lack strong doping dependence \cite{Kivelson03}. The 0.4 eV activation energy could then represent the strength of pinning of stripe-like order to antiferromagnetic layers. Such pinning of the overall anisotropic pattern would not require local pinning of many stripes and thus would be consistent with the absence of elastic neutron scattering at these temperatures. At any rate, the noise demonstrates a connection between the surprising magnetic memories and the electronic 
correlations appearing in transport effects.

We thank Hans Hilgenkamp for untwinned film and Steven Kivelson and Philip Phillips for insightful discussions. This work was supported by the U.S. DOE, Division of Materials Sciences under grants DE-FG02-07ER46453 (DSC, DVH and EHF) and DE-FG02-08ER46533 (MH and TRL), the NSF under grant DMR-0605726 (VO and MBW), and through the Frederick Seitz Material Research Laboratory at the University of Illinois at Urbana-Champaign.

\end{document}